\documentclass[review]{elsarticle}

\usepackage{lineno,hyperref}
\usepackage{amsfonts}
\usepackage{epstopdf}
\usepackage[export]{adjustbox}
\usepackage{float}
\usepackage{graphicx}
\usepackage{subfigure}
\usepackage{color}
\usepackage{bm}
\usepackage{textcomp} 
\usepackage{enumitem}
\hypersetup{
	colorlinks=true,
	urlcolor= blue,
	citecolor=blue,
	linkcolor= blue,
	bookmarksopen=false,
}
\modulolinenumbers[5]

\journal{journal of Physica E}

\begin{document}

\begin{frontmatter}

\title{Electronic and magnetic properties of honeycomb zigzag nanoribbons in the in-plane transverse electric field using Kane-Mele-Hubbard model}

\author{J. Ghorbani}
\address{Department of Physics, Qom Branch, Islamic Azad University, Qom, Iran}
\author{M. Ghaffarian}
\address{Department of Physics, University of Qom, Qom, Iran}
\ead{m.ghaffarian@yahoo.com}

\begin{abstract}
Using the Kane-Mele-Hubbard model in the unrestricted mean field approximation, the effect of spin-orbit coupling, as an intrinsic parameter, and an in-plane transverse electric field, as an external parameter, on the electronic and magnetic properties of honeycomb zigzag nanoribbons are investigated in the presence of electron-electron interaction. Our calculations show that each of these parameters has significant effects on the physical properties of nanoribbons, and each of them independently transitions the nanoribbon from a magnetic to non-magnetic state. The process of change in some aspect of physical properties, such as symmetry breaking, separation of spin up and spin down energy bands, reduction of magnetic order, change of electric dipole moment and spin current of nanoribbons, which are due to change of these two parameters are investigated. we will see that spin-orbit coupling in the competition with the transverse electric field determines the basic physical properties of the nanoribbon. This research, can provide a better understanding of two types of topological and non-topological transitions. In addition, the separation of spin-up and spin-down energy bands and their tuning by these parameters can be considered as a candidate for use in spintronic instruments.
\end{abstract}


\end{frontmatter}


\section{\label{sec:Intro}Introduction}

Graphene and graphene-like zigzag nanoribbons, because of their honeycomb structures and proper edges, show interesting magnetic and electronic properties which make them candidates for research in the field of condensed matter physics ~\cite{son2006half,chen2017width,xu2010robust,ghader2019asymmetric,luo2021full}. The edge states in this system, which is localized in zigzag edges in the tight-binding model, lies precisely at the Fermi energy, at half-filling~\cite{nakada1996edge,wakabayashi2010electronic}. The large density of edge state results in a stoner instability that leads to ferromagnetic order~\cite{stoner1938collective}. It is found that electron-electron interaction, in the mean field Hubbard model~\cite{fujita1996peculiar,carvalho2014edge,perez2012spin} as well as density functional calculation~\cite{kusakabe2003magnetic,moon2016investigations} and dynamical mean field Hubbard model~\cite{raczkowski2020hubbard}, results in the ferromagnetic order in the edge of the nanoribbon and opens a gap in band energy. 

Another remarkable electronic phase of honeycomb lattice is the topological insulator phase. In 2005, Kane and Mele predict that spin-orbit coupling leads graphene to a new electronic phase which is called the quantum spin Hall regime~\cite{kane2005quantum,kane2005z}. Using the Haldane model~\cite{haldane1988model}, they add spin-orbit interaction terms to the simple Tight- Binding Hamiltonian of graphene and calculated the band structures of the zigzag nanoribbon. It is seen that a gap opens in Fermi energy in which there are two spin filtered channels, that are protected by time-reversal symmetry. So, the spin-up and spin-down electrons could have a rotational motion in the edges of nanoribbon in opposite directions. Also, the Kane-Mele model could describe the electronic properties of other two-dimensional honeycomb lattices such as scilicene~\cite{geissler2013group,mohammadi2021topological}, germanene~\cite{si2014functionalized,azizi2021effects}, stanene~\cite{modarresi2016topological}, the double layer perovskite iridates~\cite{shitade2009quantum} and metal-organic frameworks~\cite{wang2013organic}. 

The effect of magnetic ordering on the Quantum spin Hall regime in zigzag nanoribbons was studied by adding electron-electron interaction in the frame of mean-field Hubbard term to the Kane-Mele Hamiltonian by some authors~\cite{soriano2010spontaneous,gosalbez2012spin,autes2013engineering}. They extracted a phase diagram showing that zigzag nanoribbons, depending on the relative relationship of spin-orbit coupling,  \( t_{so} \) , and Hubbard coulomb interaction, U, could have three phases of ferromagnetic, half metal, and topological insulator. They indicated that, in the weaker spin-orbit coupling rather than Coulomb interaction, band structures of nanoribbons become asymmetric. 

In this paper, like the above researchs, we also study the effect of spin-orbit interaction on the electronic and magnetic properties of honeycomb zigzag nanoribbons in the presence of electron-electron interaction, but using the transport viewpoint of the quantum spin Hall effect, which is introduced in the recent paper~\cite{baradaran2020bias}, the continuous change of topological phase of these nanoribbons is investigated. we will see that  the asymmetry of the band structure is a result of the conflict of spin-polarized current and ferromagnetic pinned electrons in the edges of ribbons. It will be seen that a smooth and continuous transition from magnetic to non-magnetic states occurs with increasing spin-orbit intensity. We also show that another transition from magnetic to non-magnetic states, for zigzag nanoribbons, occurs through an in-plane transverse electric field. By applying a transverse electric field, the degeneracy of spin-up and spin-down energy band breaks and magnetic moment in the edges of nanoribbons decrease until in a critical electric field a transition to a nonmagnetic state takes place. The phenomena resulting from these two types of transitions are discussed in detail separately. In the next step, we simultaneously considered the effect of the spin-orbit coupling and the transverse electric field on the nanoribbons. The effect of these two parameters on magnetic ordering in edges and variations of band structures are discussed.

\section{\label{sec:model} Model and formalism}

\begin{figure}[!t]
	\centering
	\includegraphics[width=0.8\textwidth]{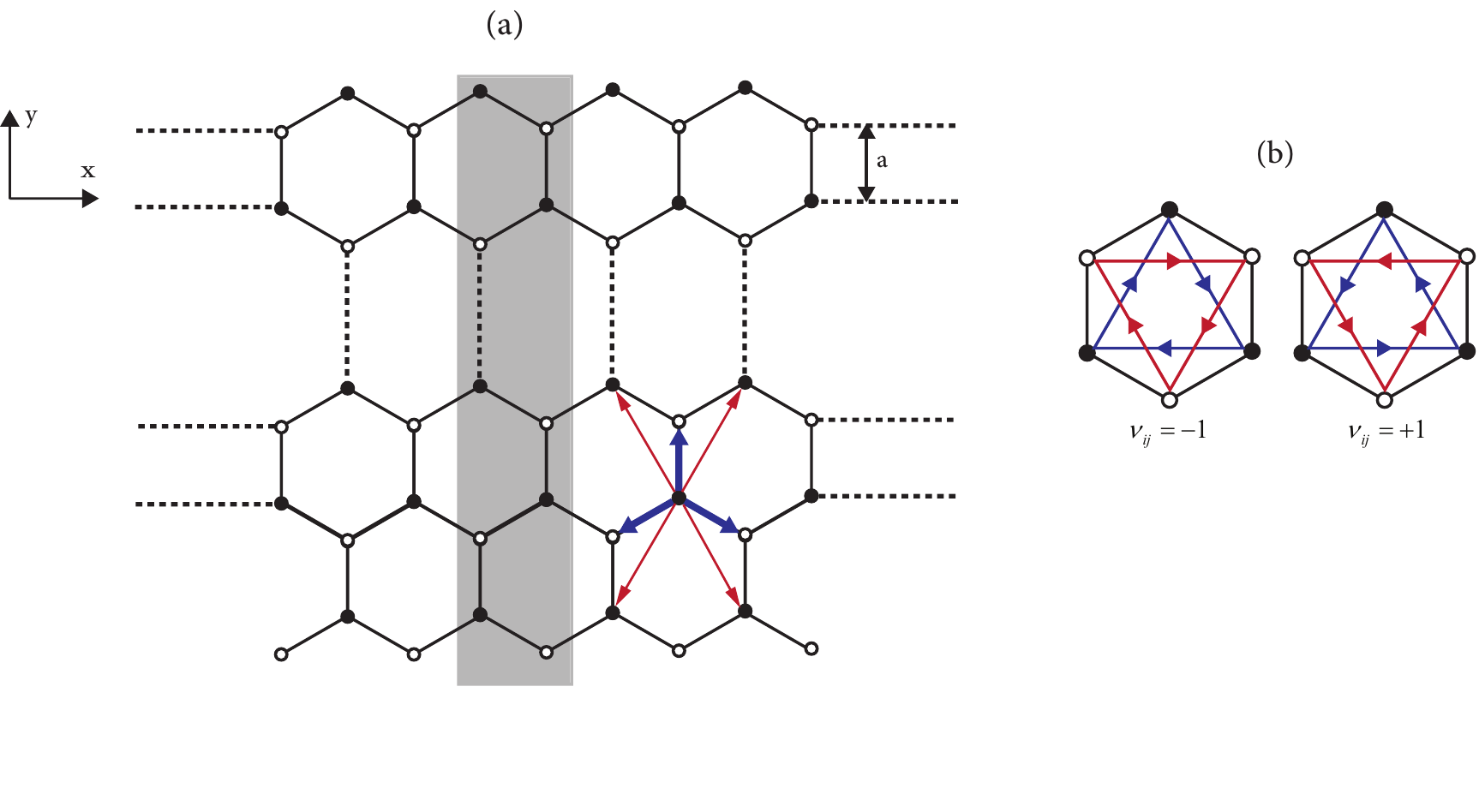}	
	\caption{\label{figure1}(color online).(a)Honeycomb zigzag nanoribbon which are infinite in the $x$-direction and finite in the $y$-direction. The origin of the coordinate system is in the middle of nanoribbon width. The shaded rectangle shows a primitive cell of the nanoribbon. (b) Arrows on second neighbor bond which shows the Haldane factor for clockwise and anticlockwise direction.  }
\end{figure}

Honeycomb zigzag nanoribbons are considered as a quasi one-dimensional system which is infinite in the $x$-direction and finite in the $y$-direction (Figure.~\ref{figure1}(a)).
The model used to study these nanoribbons, which includes spin-orbit interaction as a topological insulator parameter and Coulomb interaction, is the one orbital Kane-Mele-Hubbard Hamiltonian at half-filling, with the additional term, that considers the effect of an in-plane transverse electric field on electronic properties of zigzag nanoribbons.

\begin{equation}
\label{eqn:1}%
\vspace{\baselineskip}
\ H= \sum _{<ij> \sigma }^{}t~C_{i \sigma }^{+}C_{j \sigma }+i \sum _{ \ll ij \gg  \sigma }^{}t_{so} \sigma _{z} \nu _{ij}C_{i \sigma }^{+}C_{j \sigma }+H_{int}+H_{E} \
\setlength{\parskip}{0.0pt}
\end{equation} 

Where  \( C_{i \sigma }^{+} \)  and  \( C_{i \sigma } \)   are, respectively the creation and annihilation operators of   \(  \pi  \)-electron at $i$'th site with spin  \(  \sigma  \) .  \( <ij>and \ll ij \gg   \) indicate summation over the nearest and next-nearest neighbor, respectively. t is hopping integral while  \( t_{so} \)  is the strength of spin-orbit coupling.  \(  \nu _{ij}= \pm 1 \)  is Haldane factor for clockwise or anticlockwise second neighbor hopping (Figure.~\ref{figure1}(b)).  \(  \sigma _{z} \)  is the corresponding Pauli matrix describing the spin of the electron.
We use the Hubbard term,  \( H_{Hub}=U \sum _{i}^{}C_{i \uparrow }^{+}C_{i \uparrow }C_{i \downarrow }^{+}C_{i \downarrow } \) , to include electron-electron interaction in the mean-field approximation

\begin{equation}
\label{eqn:2}%
 \ H_{int}=U \sum _{i}^{}C_{i \uparrow }^{+}C_{i \uparrow }<C_{i \downarrow }^{+}C_{i \downarrow }>+<C_{i \uparrow }^{+}C_{i \uparrow }>C_{i \downarrow }^{+}C_{i \downarrow }-<C_{i \uparrow }^{+}C_{i \uparrow }><C_{i \downarrow }^{+}C_{i \downarrow }> \
\end{equation}

Where The quantity \textit{U }is the strength of onsite interaction.  \( <C_{i \uparrow }^{+}C_{i \uparrow }> \)  and  \( <C_{i \downarrow }^{+}C_{i \downarrow }> \)  are the expectation values of spin up and down electron densities at site \textit{i }, respectively, which must be determined self-consistently.
The last term, which is due to the coupling of the  \(  \pi  \)-electrons to the external in-plane transverse electric field, \textbf{E},  \( H_{E}=eE \cdot  \sum _{i \sigma }^{}r_{i} \)   \( C_{i \sigma }^{+}C_{i \sigma } \)  ,where \textit{e }is the charge of the electron, \textbf{E }is the in-plane transverse electric field and  \( r_{i} \) \textit{ }is the expectation value of the electron position operator, \textbf{r}, with respect to the Wannier state  \(  \vert  \varphi _{i}> \) , which is equal to the atom position,  \( r_{i}=< \varphi _{i} \vert r \vert  \varphi _{i}> \) . 


\section{\label{sec:result}Result and discussion}
Using the Kane-Mele-Hubbard model and unrestricted mean-field approximation, which was introduced in the previous section, the effect of an in-plane external transverse electric field and spin-orbit interaction on interacting honeycomb zigzag nanoribbons are considered.

we have divided this section into three subsections which in the first subsection, we will see that how the spin-orbit interaction could change the electronic structure of nanoribbons in the absence of any external field. The effect of transverse electric field on the band energy and edge magnetic state of nanoribbons with no spin-orbit interaction is studied in the second subsection. Finally in the last subsection, We investigate the simultaneous effect of transverse electric field and spin-orbit interaction on the electronic and magnetic properties of nanoribbons.

Our calculations have been performed for zigzag nanoribbons with 100 atoms in a primitive cell. We consider the intensity of onsite interaction, U, to be a constant value, $U=t$. All quantities are represented in the energy dimension in terms of the hopping integral, t, and in the space dimension in terms of the distance between two nearest neighbor atoms, a.

\subsection{Effect of spin-orbit interaction on nanoribbons in the absence of transverse electric field}
One of the characteristics of honeycomb zigzag nanoribbons is the presence of magnetic order at the edge of the nanoribbons. This magnetic order is well explained using the Hubbard model in the unrestricted mean-field approximation. In this model, the local density distribution of spin-up and spin-down electrons can be investigated by calculating their contribution at each site in the primitive cell. on the $i$’th site this is
$\rho_{\uparrow\downarrow}(i)= 1/N \sum_{n,k}^{\prime}|\psi_{\uparrow\downarrow}(i)|^{2} $
,which
$ \Psi_{n,k}^{\uparrow\downarrow} (i)$
is the $i$’th element of spin-up or spin-down eigenvector of $k$’th orbital from the $n$’th band. The sign of prime shows that the sum is over the occupied orbitals and N is the number of the primitive cell along with the ribbon. Also we assume the local spin density of electrons is $\rho_{s}(i)=\rho_{\uparrow}(i)-\rho_{\downarrow}(i)$. In our calculation, which is due to our initial condition in the self-consistent calculation, accumulation of spin-up electrons density in the upper edge and spin-down electrons density in the lower edge of nanoribbons are the cause of magnetic properties of this type of nanoribbons.

\begin{figure}[!t]
	\centering
	\includegraphics[width=1.0\textwidth]{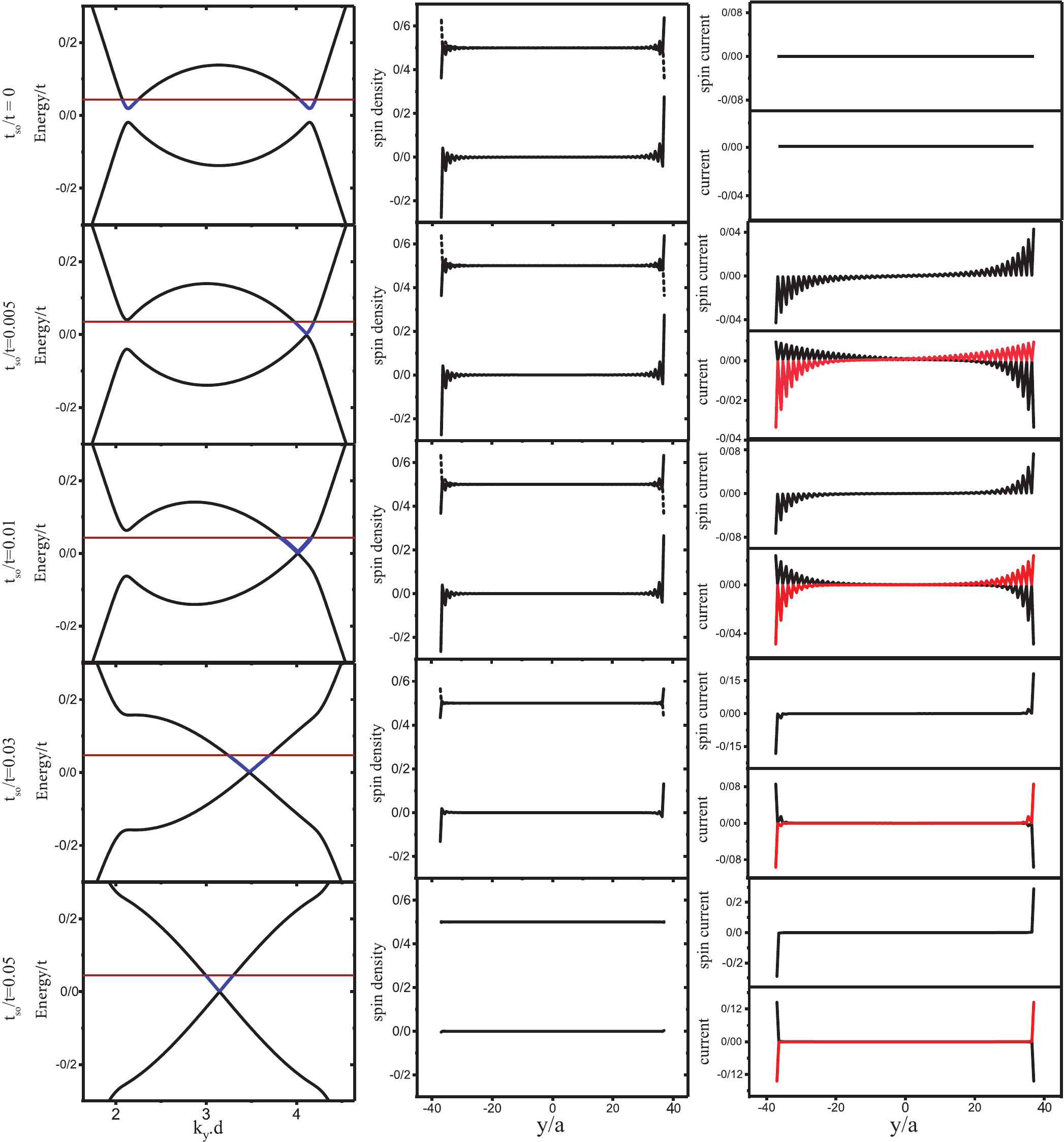}
	\caption{\label{figure2}(color online). Band structures, local spin density and spin current of honeycomb zigzag nanoribbons for different spin-orbit intensities are shown in the first, second and third columns, respectively. In the band structures diagrams, the straight line indicates the Fermi energy level when the gate voltage is applied. The filled part of the conduction band, which is below this Fermi level, is used to calculate the spin current shown in the third column. In each diagram of the second column, the lower curve represents the local spin density and the upper curves represent the local spin-up (solid line) and spin-down (dashed line) density of the  electrons. In each diagram of the third column, the upper curve represents the spin current and the lower curves represent the spin-up (red) and spin-down (black) electron current.}	
\end{figure} 

By adding the Kane-Mele term to the Hubbard Hamiltonian, the spin-orbit effect on the electronic and magnetic properties of nanoribbons can also be considered. It is seen that, in the weak spin-orbit coupling, the electronic properties of zigzag nanoribbons do not change considerably and the magnetic state of the ribbons remains the same. Some authors showed that in the strong of spin-orbit coupling, phase transition takes place from ordinary state to topological insulator state~\cite{soriano2010spontaneous,gosalbez2012spin,autes2013engineering}. We extract their results with a more detailed explanation. In Figure.~\ref{figure2} we see that, for the nanoribbons with zero spin-orbit coupling,there are two gaps in the band structures of  nanoribbons in the vicinity of two Dirac points on both sides of the Brillouin zone boundary. With consideration of spin-orbit coupling, we see that, one gap decrease and another gap increase. Therefore, the band structure symmetry breaks. By increasing spin-orbit strength, the band structure becomes more asymmetric and the single gap gradually closes (The first column of Figure.~\ref{figure2}). This state is named half metal. As the spin-orbit strength increases again, the closed gap becomes closer to the Brillouin zone boundary and the band structure gradually becomes symmetric and linear in the vicinity of Fermi energy, so nanoribbon goes to a topological insulator state.

There is a basic question. Why in the weak spin-orbit coupling, band structures become asymmetric. This answer is due to the magnetic state of zigzag nanoribbons. We simultaneously study the magnetic state of the nanoribbon and the spin current due to an energy band that is partially filled. To calculate spin current, as in previous research~\cite{baradaran2020bias}, we assume that the nanoribbons are exposed to the gate voltage so that the conduction band becomes partially filled. So, local spin-up and spin-down current, $J_{}(\vec{R})=\sum_{k}e\nu_{\uparrow(\downarrow)}(k)|\Psi_{\uparrow(\downarrow)k}(\vec{R})|^{2} $ and then local spin current $J_{s}=J_{\uparrow}-J_{\downarrow}$ are calculated. Which, the sum is over the occupied state of partially filled band energy, \textit{e} is the electron charge, $\nu_{\uparrow(\downarrow)}(k)=\frac{1}{h} \frac{\partial.pdfilon_{\uparrow(\downarrow)}(k)}{\partial k}$ is the velocity of spin-up (spin-down) electron in \textit{k} state and $\Psi_{\uparrow(\downarrow)k}(\vec{R})$ is the corresponding eigenvector in site $\vec{R}$.  The diagrams of band structures, local spin density, and local spin current of partially filled band energy for zigzag nanoribbons with various spin-orbit strength is depicted in the first, second and third columns of Figure.~\ref{figure2}, respectively. As we can see, for the weaker spin-orbit coupling, there is local spin density in the edges of the nanoribbon. As we know, spin-orbit coupling is a parameter that creates a spin current in the edges of the nanoribbon. But when its strength is not enough, it cannot destroy local spin density. So, a conflict occurs between local spin density and spin current which leads to an asymmetry of the energy band. In our calculations, the local spin-up (spin-down) density is in the upper (lower) edge of the nanoribbon. So, when we calculate local spin-up and spin-down current of partially filled band energy, we see a rotational spin current, even in the weak spin-orbit coupling, so that spin-up (spin-down) current is localized in the lower (upper) edge but is distributed in the upper (lower) half of nanoribbon. This is because of the Pauli Exclusion Principle which the pinned local spin-up (spin-down) does not permit to spin-up (spin-down) current to flow from the upper (lower) edge and exclude them toward the center of nanoribbon (Diagram of the third column of the second and third rows of Figure.~\ref{figure2}). We call this the mixed state, in which there is both a spin current and a magnetic order at the edge of the nanoribbon.  In this state, there is time-reversal symmetry with respect to the k-point in which lower gap is located but the inversion symmetry is broken. In fact, it is this inversion asymmetry in the energy band that explains the asymmetric rotation of the spin-up and spin-down currents, or it can be said that the existence of asymmetric rotation of the spin-up and spin-down currents causes inversion asymmetry in the energy band.

As the spin-orbit strength increase, the magnetic state becomes weaker and spin-up (spin-down) current flows closer to the upper (lower)  edge of nanoribbon until, in the strong spin-orbit strength, the magnetic state vanishes and they become symmetric completely.

\subsection{Nanoribbons under transverse electric field without spin-Orbit interaction}

 \begin{figure}[!t]
	\centering
	\includegraphics[width=0.95\textwidth]{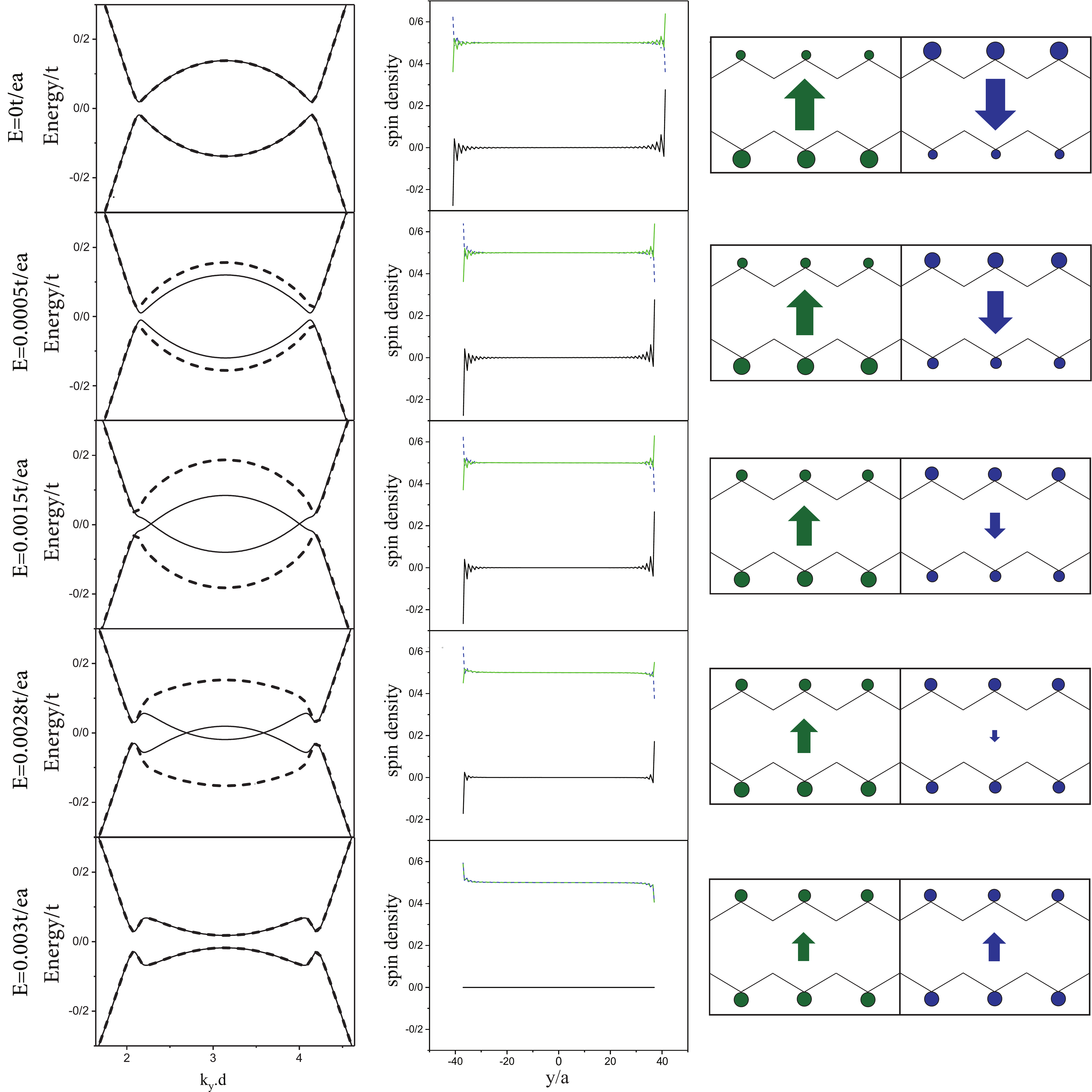}
	\caption{\label{figure3}(color online). Band structures, local spin density, and schematic representation of the electric dipole moment of spin-up and spin-down electrons of a honeycomb zigzag nanoribbon, without spin-orbit coupling, for different transverse electric fields, are shown in the first, second, and third columns, respectively. The spin-up and spin-down energy curves are shown in the band structure diagrams with a solid line and a dashed line, respectively. In each diagram of the second column, the lower curve represents the local spin density and the upper curves represent the local spin-up (solid line) and spin-down (dashed line) density of the  electrons. In the third column, in each row, the local edge distribution and electric dipole moment for the spin-up electrons in the right diagram with blue circles and arrow, respectively, and for the spin-down electrons in the left diagram with the green circles and arrow, respectively, are shown qualitatively and schematically.}	
\end{figure} 
 
The band structures of a honeycomb zigzag nanoribbon, in various transverse electric fields, are calculated and depicted in Figure.~\ref{figure3}. In the zero electric field, the spin-up and spin-down band structures are the same, and by applying a transverse electric field, the spin-up and spin-down band structures near the Fermi energy separate from each other.
The first question arises is that why the spin-up and spin-down bands separate in the effect of the transverse electric field. This is due to the non-uniform and different distribution of spin-up and spin-down electron densities along the width of nanoribbon in the zero electric field which gives rise to interesting electric and magnetic properties. One of these features, as mentioned in the previous subsection, is the creation of a net local spin density that results in a magnetic order at the edges of the nanoribbon.

 \begin{table}[!b]
	\begin{center}
		\caption{\label{tab:1}Variation of transverse spin-up, spin-down and total electric dipole moment (EDM) of honeycomb zigzag nanoribbons in some transverse electric field. Dimension of electric field and EDM is $ t/ea $ and $ ea $, respectively  }
		\begin{tabular}{c c c c c c c c}
			\hline \hline
			Transverse& 0.0000 &0.0005 &0.0010 &0.0015 &0.0020  &0.0025  &0.0030  \\
			electric field          &    &     &  &  & & &  \\ \hline
			Spin-up EDM       & -11.60&-11.36 &-10.54 &-8.58&-5.20&-2.65 & 11.96 \\ 
			Spin-down EDM      & 11.60 &11.76  &11.88  &11.98& 12.06&12.14 & 11.96   \\ 
			Total EDM          & 0     & 0.4   &1.34   &3.4&6.86  & 9.49   & 23.92   \\ 
			\hline \hline
		\end{tabular}
	\end{center}
\end{table}

 Another physical property, which results from the non-uniform distribution of spin-up and spin-down electrons densities, is the existence of permanent spin-up and spin-down electric dipole moment (EDM) in the direction of the width of the nanoribbon. In our calculations, the permanent spin-down EDM is in the y-direction while the magnitude of spin-up EDM is the same as the magnitude of spin-down EDM but is in the opposite y-direction, so the total EDM is zero (Table.~\ref{tab:1}). This phenomenon shows that the response of spin-up and spin-down electrons to the transverse electric field is different. For example, if we apply an external electric field in the y-direction the spin-down EDM is in the direction and the spin-up EDM is in the opposite direction of the external electric field. Thus, spin-up electrons are more affected by the electric field than spin-down electron. With the applying and increase of the transverse electric field, spin-up EDM changes very quickly while the change of spin-down EDM is much slower (Table.~\ref{tab:1} and The third column of Figure.~\ref{figure3}). So the electric susceptibility for spin-up electrons is much more than spin-down electrons. 
 
 It seems that, because of the opposite direction of spin-up and spin-down EDM, the contribution of spin-up and spin-down electrons in the total energy are different from each other in the external transverse electric field. So the changes of their band structures in the effect of the external electric field are different too. As shown in the first column of Figure.~\ref{figure3}, by applying a transverse electric field, the spin-up and spin-down band structures near the Fermi energy separate, and the energy gap of the spin-down bands increase, while the energy gap of the spin-up band decrease and in a given electric field this gap closes. This behavior of the band structures in the effect of the transverse electric field is so interesting because the separation of spin-up and spin-down band structures is usable for spin-filtered devices. Especially when the gap between spin-up bands is closing while there is a significant gap between spin-down bands.  So, this device has conductor behavior for the spin-up electrons and semiconductor behavior for the spin-down electrons.

The evolution of magnetic properties of honeycomb zigzag nanoribbons in the effect of the transverse electric field is shown in the second column of Figure.~\ref{figure3}. As the figure shows, Spin-up electrons are more concentrated at the upper edge and spin-down electrons are more concentrated at the lower edge.
 By applying a transverse electric field in the y-direction, the spin-up and spin-down electrons are pulled from the upper edge to the lower edge, but the transfer rate for spin-up electrons (whose EDM direction is opposite to the electric field direction) is much greater than the transfer rate for spin-down electrons (whose EDM direction is in the electric field direction). 
 Therefore, the spin density of electrons generally decreases with the increasing transverse electric field, and the magnetic property is weakened.

 \begin{figure}[t]
 	\centering
 	\includegraphics[width=6.7cm]{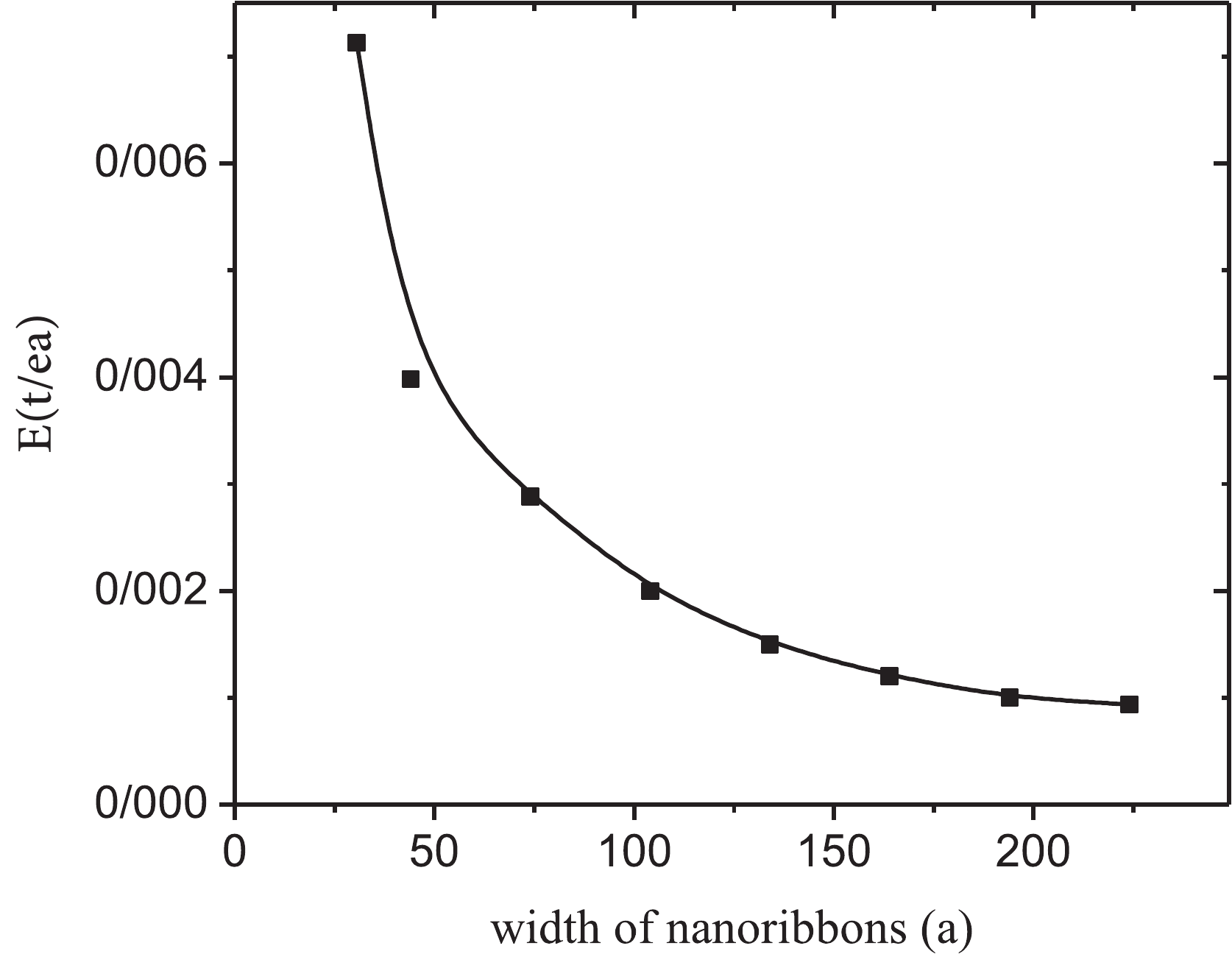}
 	\caption{\label{figure4}Critical transverse electric field for several honeycomb zigzag nanoribbons of different widths. }
 \end{figure}

  As before explained, with the increasing electric field, the gap of spin-down energy bands increases and, the gap of spin-up energy bands close. In this case, two crossing points are created for the spin-up electrons energy bands around the Dirac points, as the electric field increases, these two points move towards the first Brillouin zone boundary, Until in a particular electric field, a phase transition, from magnetic to nonmagnetic state, occurs (The last row of Figure.~\ref{figure3}). In this state, spin-up and spin-down band structures and their spatial eigenvectors become identical. the Density of spin-up and spin-down electrons on each site become equal, so the spin density, on each site, vanishes. According to our calculations, wider ribbons are more sensitive to the transverse electric field and the transition phase takes place in a weaker transverse electric field (Figure.~\ref{figure4}).

\subsection{Spin-orbit effect on nanoribbons under transverse electric field}
In the two last subsections, we see that two physical parameters, an in-plan transverse electric field, and spin-orbit coupling, could change the magnetic state of honeycomb zigzag nanoribbons. With increasing of each of which these parameters, individually, the magnetic ordering of edges of nanoribbons decrease until they transit to a non-magnetic state.  In this subsection, we study the effect of both parameters, simultaneously, on  the electronic and magnetic properties of honeycomb zigzag nanoribbons.

\begin{figure}[!t]
	
	\centering
	
	\includegraphics[width=1.0\textwidth]{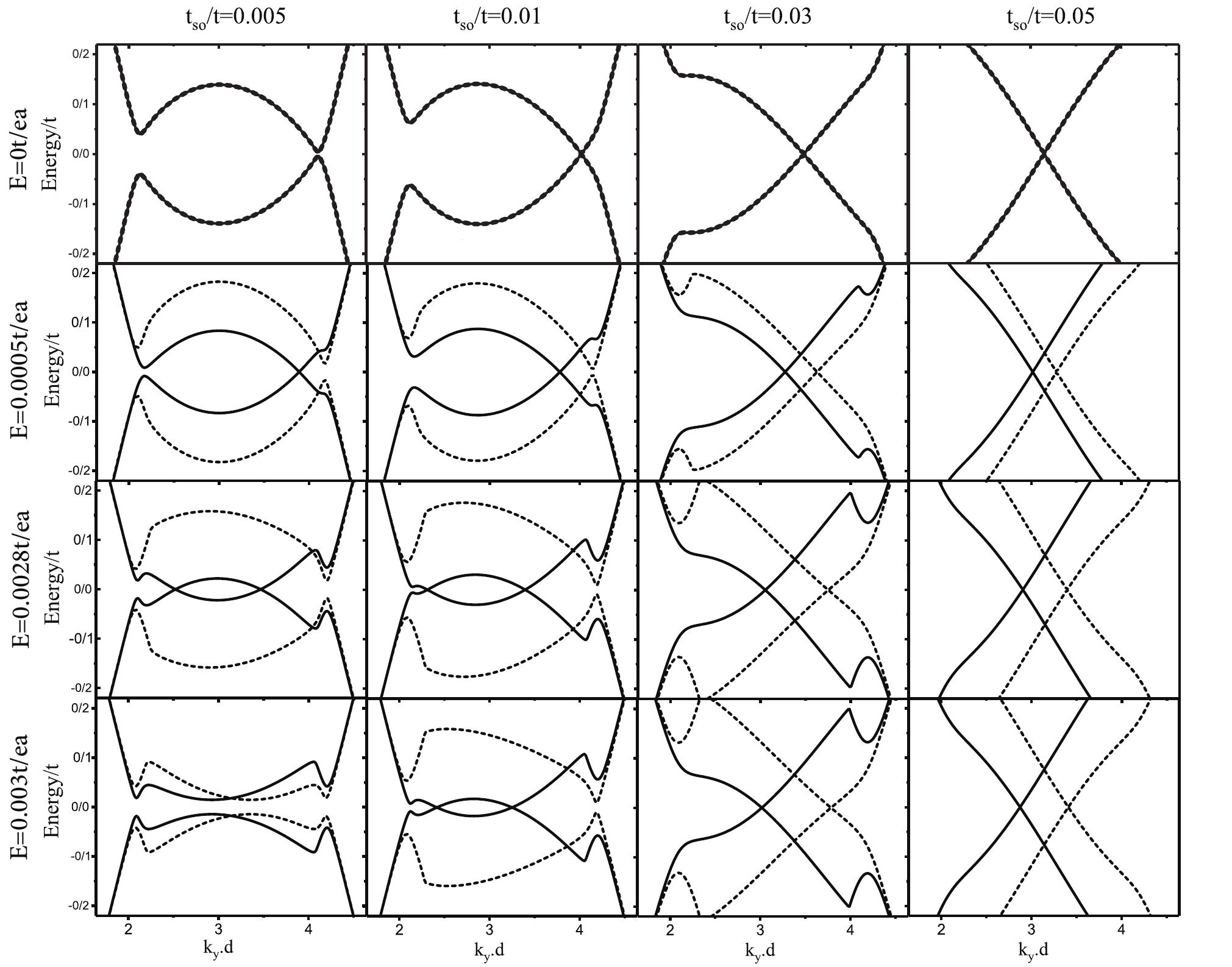}
	
	\caption{\label{figure5}Each row shows the change of the band structures of honeycomb zigzag nanoribbons with different spin-orbit intensities, which are under a constant transverse electric field. Each column shows changes in the band structure of a honeycomb zigzag nanoribbon with fixed spin-orbit intensities exposed to different transverse electric fields. In each diagram, the solid-line and dashed curves represent the spin-up and spin-down energy bands, respectively.}
\end{figure}

 In Figure.~\ref{figure5}, diagrams of the band structure of nanoribbons, in various strengths of the transverse electric field and spin-orbit coupling, are depicted. This figure illustrates the effect of a transverse electric field on nanoribbons in the mixed state. The mixed state, as we saw in the first subsection is a state in which there is both an edge spin current and an edge magnetic order in the nanoribbon. The electric field, in this case, separates the spin-up and spin-down energy bands, but this separation is both in terms of energy and in terms of wavenumber. In fact, in the mixed-phase, there is both a weak edge spin current and also the opposite spin-up and spin-down electric dipole moments due to the imbalance local density of spin-up and spin-down electrons at edges. It is shown that, the effect of transverse electric field on opposite spin-up and spin-down electric dipole moments cause the separation of energy bands in the direction of energy (subsection.2) and on the edge spin current causes the separation of bands in the direction of the wavenumber~\cite{baradaran2020bias,liu_AppPhysLett.99.2011_EFQSHE}. Therefore, the transverse electric field well indicates the simultaneous existence of these two phenomena in nanoribbons.

Each column in Figure.~\ref{figure5} shows the effect of increasing the electric field on the energy bands of a nanoribbon with a specific spin-orbit intensity. In each row, the effect of a constant transverse electric field on the energy bands of nanoribbons with different spin-orbit intensities can be seen. It is clear from the first and second columns of Figure.~\ref{figure5} that by applying a transverse electric field to nano-ribbons with weak spin-orbit intensity, the time-reversal symmetry is also broken, and as the field increases, the separation of the spin-up and spin-down energy bands and the asymmetry increase to the extent that at a critical electric field strength, the nanoribbon transitions from magnetic to non-magnetic state. In this case, there is time-reversal symmetry, but there is no inversion symmetry (The last diagram of the first column of Figure.~\ref{figure5}). It is evident from the last row of Figure.~\ref{figure5} that for nanoribbons, which are under a given transverse electric field, the transition from magnetic to the nonmagnetic state does not occur for those have stronger spin-orbit coupling. In the following description, we will see that the stronger the nanoribbon spin-orbit intensity, the greater the critical electric field.

\begin{figure}[t]
	\includegraphics[width=6.7cm]{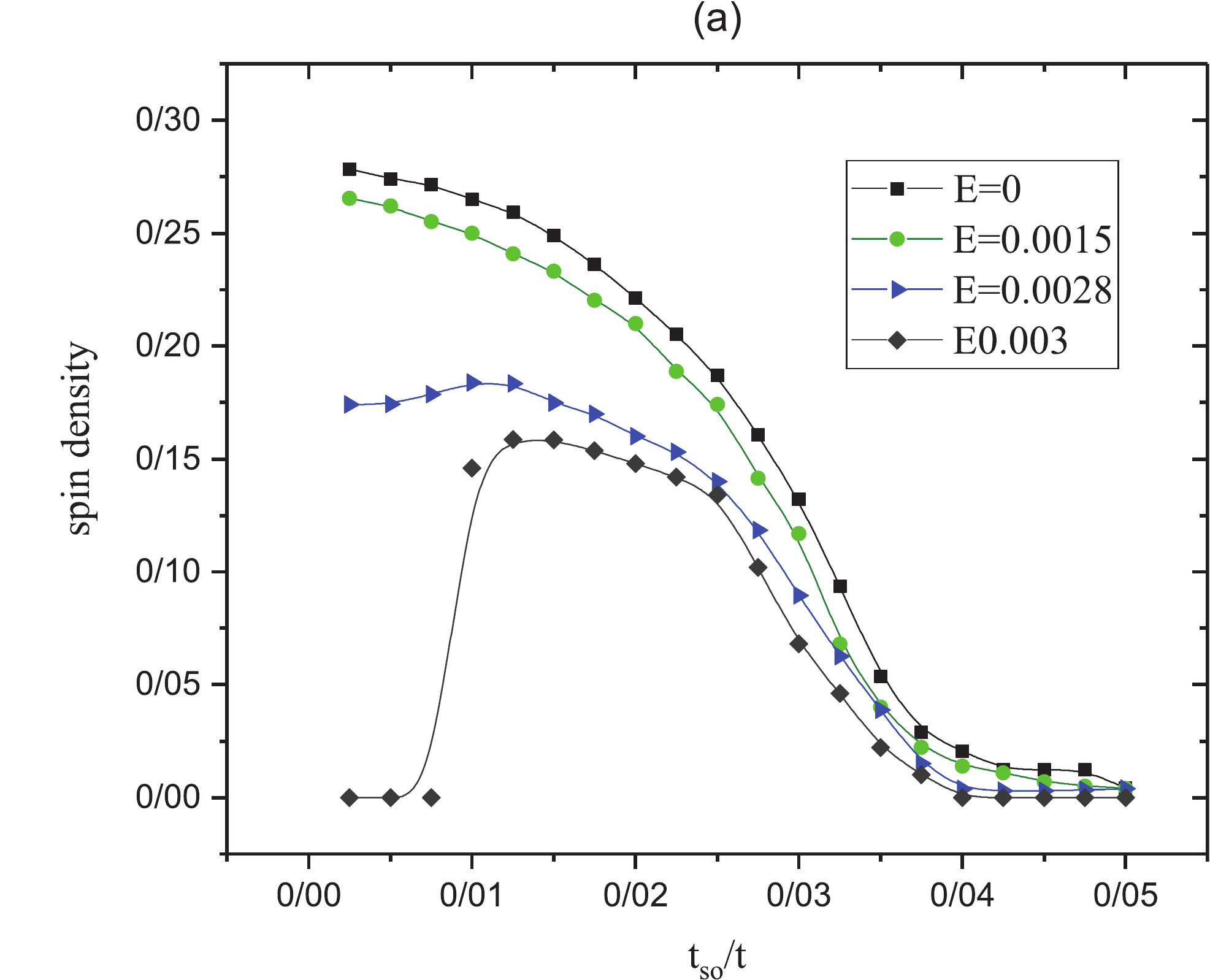}
	\includegraphics[width=6.7cm]{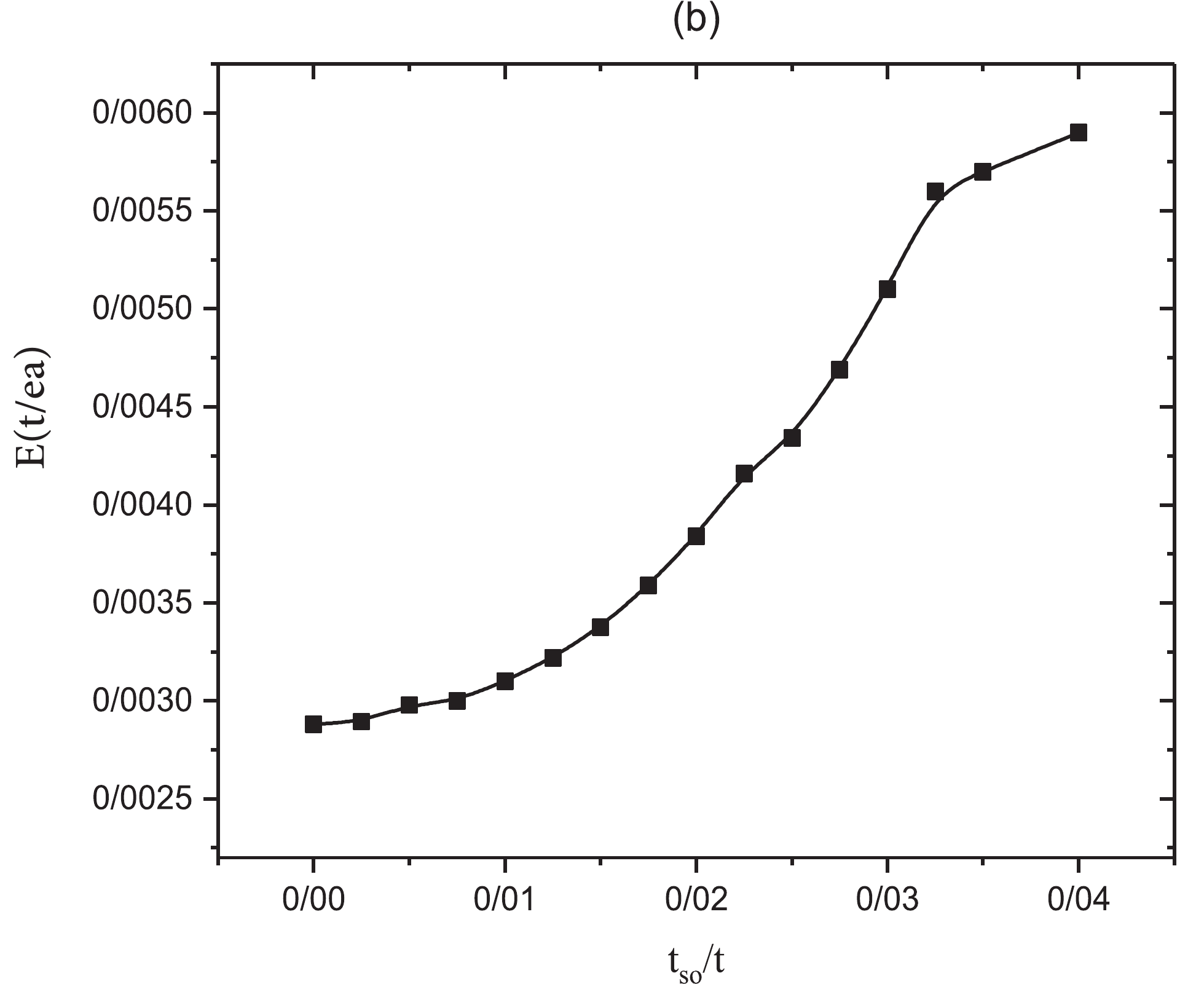}
	\caption{\label{figure6}(color online).(a) Local spin density at the edge of honeycomb zigzag nanoribbons in terms of spin-orbit intensity exposed to different electric fields. (b) Critical transverse electric field in terms of spin-orbit intensity for honeycomb zigzag nanoribbons. }
\end{figure}

The local spin density at the edge of the nanoribbon is plotted in terms of the spin-orbit coupling intensity in different transverse electric fields in Figure.~\ref{figure6}(a). As we see, in general, the local spin density at the nanoribbon edge decreases as the spin-orbit increases or the transverse electric field increases. However, there is an exception, in which the nanoribbon is in a nonmagnetic state due to the application of a transverse electric field (The last diagram of the first column of Figure.~\ref{figure5}). In this case, we see that as the spin-orbit intensity increases, the local spin density of the edges suddenly increases significantly for a specific spin-orbit intensity, $t_{so}=0$.0077t (Figure.~\ref{figure6}(a), curve due to electric field=0.003 $\frac{t}{ea}$). In fact, unlike the previous behavior, increasing the intensity of the spin-orbit increases the magnetic moment of the edge. By studying the nanoribbon energy curves (The last row of Figure.~\ref{figure5}), it can be seen that increasing the spin-orbit intensity changes the nanoribbon phase from nonmagnetic to a mixed state.

 After this transition, we see that with a gradual increase in spin-orbit intensity the local spin density of the edges gradually increases, and after reaching a maximum point it begins to decrease until, in a given spin-orbit intensity, disappears. This behavior is seen in the mixed state when the nanoribbon is under a strong transverse electric field (Figure.~\ref{figure6}(a), curve due to electric field=0.0028 $\frac{t}{ea}$). The reason for this behavior is due to the different effects that the transverse electric field and the spin-orbit interaction have on the electronic properties of the nanoribbon. As explained in subsection 2, the transverse electric field reduces the spin density of the edge due to the imbalance of the local density of the spin-up and spin-down electrons at the nanoribbon edge, while the spin-orbit interaction changes the topological phase of the nanoribbon from trivial to nontrivial. This phase transition, which takes place slowly, gradually reduces the local spin density of the edge and creates a spin current at the edge of the nanoribbon. Therefore, this effect reduces the imbalance in the density of spin-up and spin-down electrons at the edge by gradually changing the topological phase. As a result, the spin-up and spin-down transverse electric dipoles are attenuated, and the effect of the transverse electric field on the electronic properties of the nanoribbon is reduced, so the phase transition from magnetic to non-magnetic occurs in stronger electric fields. This phenomenon is well illustrated in Figure.~\ref{figure6}(b), which shows the critical transverse electric field for nanoribbons with different spin-orbit intensities. In this diagram, we see that as the intensity of the spin-orbit increases, the critical transverse electric field increases too.
 
 Examining the transition of nanoribbons from mixed to non-magnetic state due to transverse electric field, we see that this transition occurs suddenly in nanoribbons with weak spin-orbit coupling and the electronic properties of the system change suddenly. But for those with stronger spin-orbit coupling, the transition is almost smooth and the electronic properties of the system do not change. In these cases, the nanoribbons are in the topological insulator phase. In general, if we assume the spin-orbit interaction as a variable quantity, according to the mentioned results and examining each row of Figure 3, we see that the band structure of nanoribbons, independent of the electric field intensity applied to them, with increasing spin intensity, Near the Fermi level, become linear and creates a spin current that is affected by the electric field. Even in the last row of this shape, where the nanoribbon is in a non-magnetic state due to the transverse electric field, with increasing spin-orbit intensity, it first returns to the mixed state and then due to the changing topological nature of the system, the nanoribon goes to a non-magnetic state.
 
 \section{Summary and conclusions}
We used the Kane-Mele-Hubbard model in the unrestricted mean field approximation to calculate some electronic and magnetic properties of honeycomb zigzag nanoribbons which are under a transverse electric field. We investigated the effect of  spin-orbit coupling and transverse electric field on the physical properties of these nanoribbons in the presence of electron-electron interaction which are summarized below.

The presence of a relatively weak spin-orbit coupling causes the inversion symmetry of the energy band to be lost due to the existence of a weak spin current in the presence of an edge magnetic order. Increasing the intensity of spin-orbit coupling reduces the magnetic order of the nanoribbon edges and increases the edge spin current which is due to the smooth change of the topological phase of the material from trivial to nontrivial.

Applying a transverse electric field to a nanoribbon without spin-orbit coupling separates the spin-up and spin-down band structures in the direction of energy, as well as reducing the edge magnetic order. The cause of these phenomena was an imbalance in the local density distribution of spin-up and spin-down electrons at each atomic site. In addition, self-consistent energy band calculations showed that the nanoribbon transitions from a magnetic to a non-magnetic state in a critical electric field.

Applying a transverse electric field to nanoribbons with relatively weak spin-orbit coupling, causes the loss of time-reversal symmetry and the separation of the spin-up and spin-down band structures in terms of energy and wave number. In this case, in a critical electric field, nanoribbons transition from magnetic to non-magnetic state, but the intensity of this critical field increases with increasing spin-orbit intensity. The transverse electric field does not affect on the transition from the normal phase to the topological insulator phase due to the increase in spin-orbit intensity, and all nanoribbons, regardless of the electric field intensity applied to them, enter the topological phase at a certain spin-orbit intensity. Therefore, spin-orbit coupling in competition with transverse electric field determines the basic physical properties of nanoribbons.

Our purpose in this article is to provide a conceptual and physical discussion of the phenomena summarized above, and we hope that these discussions will encourage readers to offer other perspectives on the subject. We also think that some of these phenomena, especially the separation of spin-up and spin-down energy bands of honeycomb zigzag nanotubes by applying a transverse electric field and tuning them by changing the intensity of this field as well as changing the intensity of spin-orbit coupling, can be used for application in spin-filtered devices.

\newpage


\vspace{15pt}
\bibliography{ref}

\end{document}